# Enhancing autonomous vehicle acceptance with age and education sensitive simulation interventions: an experimental trial

Kacperski, Celina[1,2] Ulloa, Roberto[1,3] Wautelet, Jérémy[4] Vogel, Tobias[5] Kutzner, Florian[2]

[1]Konstanz University
[2]Seeburg Castle University
[3]GESIS – Leibniz Institute for the Social Sciences
[4] Inetum
[5]Darmstadt University of Applied Sciences

Corresponding author: celina.kacperski@uni-konstanz.de, Universitätstraße 10, 78457 Konstanz, Germany

**Acknowledgements**

Financial support by the European Union Horizon 2020 research and innovation programme is gratefully acknowledged (Project PAsCAL, grant N°815098).
We thank Maria Perez Ortega and Aaron Heinz for their support.

**Competing interests:** Authors declare that they have no competing interests.

Abstract

The familiarity principle posits that acceptance increases with exposure, which has previously been shown with in vivo and simulated experiences with connected and autonomous vehicles (CAVs). We investigate the impact of a simulated video-based first-person drive on CAV acceptance, as well as the impact of information customization, with a particular focus on acceptance by older individuals and those with lower education. Findings from an online experiment with N=799 German residents reveal that the simulated experience improved acceptance across response variables such as intention to use and ease of use, particularly among older individuals. However, the opportunity to customize navigation information decreased acceptance of older individuals and those with university degrees and increased acceptance for younger individuals and those with lower educational levels.

## Introduction

Connected and autonomous vehicles (CAVs) have the potential to enhance safety, accessibility, energy efficiency, land use, and affordability of transportation systems (Bissell et al., 2020; Faisal et al., 2019; Kacperski et al., 2020). However, the successful deployment of CAVs at scale necessitates that their performance aligns with societal expectations for safety, sustainability, and utility (Golbabaei et al., 2020; Kacperski et al., 2021), and that inequalities are not exacerbated (Cohn et al., 2019; Kacperski et al., 2024). As such, understanding the factors that influence public acceptance and behavioral intention to use CAVs is crucial for the widespread adoption of this transformative technology, especially for those populations that already face mobility related disadvantages, such as older citizens (Dabelko-Schoeny et al., 2021; Ozbilen et al., 2022) or those with lower socioeconomic status (SES), particularly lower educational attainments (Gamsjäger, 2015; Groot et al., 2012; Sadeghvaziri et al., 2023).

Research into CAV acceptance has been rapidly increasing since 2014, with over 100 articles published in 2021 alone (Ho et al., 2023). Multiple reviews have been conducted on the topic, showcasing that most studies are descriptive or correlational, and that acceptance varies widely across socio-demographic strata (Golbabaei et al., 2020; Ho et al., 2023; Jing et al., 2020; Kaye et al., 2021; Nordhoff, Kyriakidis, et al., 2019; Zhang et al., 2023). Of the socio-demographic factors that predict acceptance of CAVs, age and educational attainments stand out as two of the most consistent (Becker & Axhausen, 2017; Daziano et al., 2017; Golbabaei et al., 2020; Haboucha et al., 2017; Hardman et al., 2019; Huang et al., 2022; C. Lee et al., 2019): across much of the literature, older individuals have been shown to be less willing to adopt CAVs, often due to safety concerns (Shi et al., 2021; S. Wang & Zhao, 2019) and uncertainty about usage, usefulness, and usability (Hassan et al., 2021; Schoettle & Sivak, 2014); and participants with lower education have been found to show lower acceptance, often due to higher concerns regarding safety, concerns about the effort required to operate the CAVs, and lower beliefs about their sustainability (Akuh et al., 2023; Hudson et al., 2019; Shi et al., 2021; Thomas et al., 2020). The question of what increases acceptance of CAVs among these populations is important to investigate considering that these disadvantaged populations might most benefit from the introduction of CAVs if implemented in an inclusive, easy-to-use and sustainable manner (Hassan et al., 2021; Severs et al., 2022).

We conduct an experiment to study the efficacy of a simulated experience with a CAV, as a type of familiarization intervention. First, we present how familiarity theory underlies our intervention to encourage acceptance. Next, we review social-cognitive perspectives on individual differences in acceptance of novel technologies such as CAVs, including the role of interface design in shaping acceptance. Finally, we highlight challenges for policy and practice.

We contribute to the literature by providing evidence on the effectiveness of simulated CAV experiences based on social-cognitive theory such as the familiarity principle. We further the understanding of the interaction between individual differences and such interventions designed to increase CAV acceptance. While prior studies have noted the association of factors like age, gender, education, and technological affinity, our findings emphasize the effectiveness of interventions across age and educational attainments. This allows targeted recommendations for designing more equitable strategies to promote CAV adoption. We also contribute to the understanding of how interface design, specifically the option to customize informational feedback, impacts acceptance. While customization has been linked to increased usability and control in other technological contexts (N. Franke et al., 2009; Miraz et al., 2021), its role in the acceptance of CAVs remains underexplored (DeGuzman et al., 2022). Adaptive interfaces may affect users differently based on age and educational background, which could refine our understanding of how to design user-centered interfaces that are inclusive and cater to diverse population needs. Beside effects on intention to use, we study the effects on self-reported trust, ease of use and safety expectations, contributing to a better understanding of further facets of acceptance, and how they differ for individuals of various ages and educational attainments. Finally, much of prior research on CAV acceptance has relied on either observational, pre-post, or correlational designs, with limited use of randomized experimental methodologies. Our study extends this by employing an experimental approach, providing a stronger basis for causal inferences about the role of simulated exposure in shaping behavioral intentions.

## Literature Review

### Familiarity principle in the context of CAV acceptance

Acceptance of CAVs might be improved by direct experience with such vehicles. The familiarity principle posits that individuals tend to develop a preference for something merely due to being familiar with it, i.e., people exposed to something more often, tend to like it more (Bornstein, 1989; Montoya et al., 2017). This effect, also sometimes termed "mere exposure effect", has been extensively studied (Bornstein & Craver-Lemley, 2022); for technology in general, some evidence exists for example about exposure to computers and smartphones (Chen et al., 2016; Feindt & Poortvliet, 2020; McQuarrie & Iwamoto, 1990; Wicki, 2021).

In line with this principle, researchers have found increases in CAV acceptance with pre-post research, studying experiences in shared automated shuttles (Molina et al., 2021; Nordhoff, de Winter, et al., 2019; Papadima et al., 2020). Across various field trials, participants' safety concerns were reduced (Distler et al., 2018), and trust and perception of reliability and comfort were increased (Classen et al., 2021; Mason et al., 2022; Nordhoff et al., 2020; Paddeu et al., 2020; Zoellick et al., 2019). Positive effects were shown especially for participants of higher age (Nordhoff et al., 2018) and lower education (Dai et al., 2021). Single-person vehicles have been more rarely utilized in trials, but also show increases in acceptance (Detjen et al., 2021; Feys et al., 2021; Shi et al., 2021; Xu et al., 2018).

In vivo field trials are costly and can only serve a handful of participants at a time, making them unsuitable for large-scale interventions. In response, researchers can use simulations to explore effects on acceptance of CAVs, finding, in line with evidence from in vivo experiences, that simulated experiences increase trust and reduce safety concerns and anxiety (Buckley et al., 2018; Clement et al., 2022; Gold et al., 2015; Molnar et al., 2018; Park et al., 2019; Zontone et al., 2020). A direct comparison of the type of driving experience (shuttle or simulator) in older adults found they performed equally well in terms of increasing trust and perceived safety (Classen et al., 2020, 2021), indicating that there might only be minimal benefits to using in vivo experiences to increase acceptance compared to simulations. Nevertheless, none of these studies employ experimental methodology comparing interventions to a control group, with pre-post measurements are the standard methodology used. And while there are first indications that older and less educated individuals might benefit from more exposure to CAVs, there is little experimental evidence of the differential impact of experiences on acceptance overall, nor for these subgroups specifically.

## Individual differences in experience with and acceptance of autonomous vehicles

According to social-cognitive theories, familiarity fosters a more positive attitude toward the technology, a theoretical assumption grounded in a functional perspective (Schwarz et al., 2021; Zajonc, 1980): familiarity signals that an object has not posed a threat in the past, and conversely, lack of experience with an object is associated with heightened uncertainty and perceived risk, which negatively impacts attitudes (Fang et al., 2007; Song & Schwarz, 2009).

Individual differences here play a role for the effectiveness of a familiarity intervention: people vary in prior experience, whether through general knowledge about technology, or specific familiarity with autonomous features in their environment. Indeed, research with drivers that report prior experience with autonomous features suggests that they perceive autonomous technologies as less risky (Brell et al., 2019) - thus, individuals with prior knowledge or experience may benefit less from additional exposure to a technology.

There is a pre-existing relation with socio-demographics: older individuals often have less novel technology experience and higher risk aversion (Albert & Duffy, 2012; König, 2021; Mostaghel, 2016). As a result, they may be more receptive to a familiarity intervention such as the one we propose. Conversely, higher education is associated with greater technological knowledge (Falk & Needham, 2013), suggesting that individuals with higher education may have a reduced need for more exposure, i.e., additional familiarization may have a smaller effect for them. Evidence supports this idea: both age-related and educational disparities in access to vehicles with autonomous technologies have been noted in a study of over 1500 drivers in the U.S. (Louis et al., 2023), in a study of almost 3000 drivers where prevalence of in-vehicle autonomous features was found to be higher for highly educated drivers (Eby et al., 2018); and in a survey of 634 US drivers that measured understanding of advanced driver system features, younger men generally self-reported higher hands-on experience and confidence (Greenwood et al., 2022). In a study of 229 drivers mostly from the U.K, education was correlated with prior knowledge about CAVs, while for age, knowledge increased to peak in the 40s and then sharply declined (Thomas et al., 2020). Finally, in a survey of over 3500 U.S adults, demographic factors such as lower age and higher education were shown to better predict comfort with higher level autonomous vehicles (C. Lee et al., 2019).

Some research goes into further detail as to reasons for varying attitudes. In the autonomous vehicle acceptance literature, the lower acceptance of older adults seems driven by trust and safety concerns (Günthner & Proff, 2021; Thomas et al., 2020; S. Wang & Zhao, 2019), fears about age-related needs (Srour Zreik et al., 2023), uncertainty about how useful the technology can be (Günthner & Proff, 2021; Park et al., 2021), and how easily usable it is (Günthner & Proff, 2021; Hassan et al., 2021; Schoettle & Sivak, 2014). Safety concerns are also prominent for adults with lower education (Nielsen & Haustein, 2018), particularly regarding system failure (Thomas et al., 2020). Concerns about usefulness have here also been reported (Acharya & Mekker, 2024). Additionally, higher educated drivers have shown to be more positive about ease of use and expected effort, time and environmental impact of CAVs (Akuh et al., 2023; Nielsen & Haustein, 2018).

**Individual differences in acceptance of customization options in informational interfaces**

In line with existent research on individuals differences in information processing (Ackermann, 2003; Phillips & Sternthal, 1977), the type and amount of information presented to individuals in CAV dashboards could impact their acceptance. Suitable in-vehicle user interface (UI) should display sufficient and necessary information (S. Lee et al., 2020; Oliveira et al., 2018) - at Level 5, this information should be minimal, as the vehicle is expected to take all decisions in a safer manner than humans can (Khan et al., 2022). However, value might be derived from letting drivers customize the interface and the amount of information they receive (DeGuzman et al., 2022; S. Lee et al., 2020; Normark, 2015; Wintersberger et al., 2019), for example how much feedback they want about route information (von Sawitzky et al., 2019). A positive effect of a familiarity intervention may be counteracted when users feel customization options are absent, as those allow users to suit informational input to their needs and wants, and has been shown to positively affect perceived usability (McGrenere et al., 2007; Y. Wang et al., 2016), which has been linked to a sense of control (Hui & See, 2015) and more positive attitudes, including a higher intention to use customizable products (N. Franke et al., 2009). A cognitive explanatory pathway here is fluency, the ease with which information or experiences are processed (Schwarz et al., 2021). For individuals who may rely more heavily on customization to make the technology feel usable and intuitive, the absence of customization to match their needs could hinder fluency, thereby reducing the positive effects of familiarity on their intentions to use CAVs.

However, customization may also in some cases be detrimental if it increases the perceived riskiness of the autonomous system and therefore trust in its safety (Beller et al., 2013): according to Grice's conversational maxims, offering extensive customization choices may imply that the autonomous vehicle is not fully capable of handling all driving tasks on its own (Grice, 1975). Given that a strong sense of control has been argued to be the most important feature of a CAV's UI, which at Level 5 effectively turns the driver into passenger (Kun et al., 2016), users may infer that the vehicle requires their input and intervention, which could undermine perceptions of its self-driving capabilities and the user's trust in the system. Particularly older and less educated individuals have previously been found to be most concerned about safety, reliability and control (Günthner & Proff, 2021; Hassan et al., 2021; Hudson et al., 2019; Nielsen & Haustein, 2018; Schoettle & Sivak, 2014).

Finally, customization preferences might also vary due to accurate self-assessment in technology use. Older individuals may struggle with complex operating environments due to rarer experiences with modern interfaces and declining cognitive abilities (Günthner & Proff, 2021; Hassan et al., 2021; Srour Zreik et al., 2023). In contrast, higher education, often linked to greater cognitive performance (Schneeweis et al., 2014), can facilitate adaptation to complex interfaces (Deary et al., 2009). Therefore, older individuals might be more likely to prefer simpler or more familiar environments, while more highly educated individuals may appreciate more complex setups.

These arguments highlight the importance of balancing familiarity-building interventions with interface designs that promote user control and adaptability, particularly suited towards matching the needs and wants of specific subgroups of potential users.

**Policy challenges: towards more inclusive autonomous mobility**

To reiterate, ensuring the widespread and equitable adoption of autonomous vehicles is crucial, as they hold the potential to significantly improve mobility for disadvantaged populations. We choose to focus on age and education above other socio-demographic factors, as older adults and those with lower socioeconomic status, including lower educational attainment, often face greater mobility challenges and barriers (Dabelko-Schoeny et al., 2021; Ozbilen et al., 2022). Research highlights that mobility systems must be adapted to meet the unique needs of these groups: for instance, older adults often face physical and cognitive limitations that require tailored solutions, such as user-friendly interfaces and accessible vehicle designs (Coughlin, 2001; Webber et al., 2010). Similarly, individuals with lower education attainments may prioritize affordability and accessibility when evaluating transportation options (Pyrialakou et al., 2016). Autonomous vehicles, especially also integrated as part of the public transport system, could help address these gaps by providing affordable, accessible, and reliable transportation options (Hassan et al., 2021; Severs et al., 2022). However, research is needed to understand how to design autonomous vehicle experiences that best cater to the unique needs and concerns of these target groups to facilitate their acceptance and adoption. Studies have also emphasized the importance of involving disadvantaged communities, including the elderly and visually impaired, in the design and planning of mobility systems to ensure their needs are adequately met (Brinkley et al., 2020; Kacperski et al., 2024). By addressing the barriers faced by disadvantaged populations, policymakers can help maximizing the transformative potential of autonomous vehicles.

## Research questions and hypotheses

Our study investigates whether experiencing a simulated video-based first-person autonomous drive (in a Level 5 personal vehicle) can change behavioral intention to use CAVs (BI), specifically for individuals that have been shown to have lower acceptance of CAVs such as older people and people with lower education (**RQ1**). We also investigate how the option to customize information would affect individuals' behavioral intention to use CAVs (BI), specifically whether it will increase acceptance for older people and those with lower education (**RQ2**).

We propose the following hypotheses for **RQ1**:

**H1a:** The experience of a CAV video simulation (vs control) will increase BI, and **H1b**: will do so more strongly for individuals of higher ages and **H1c**: will do so more strongly for individuals with lower educational attainments.

We propose the following hypotheses for **RQ2**:

**H2a**: Information about customization (vs control) will increase BI, and **H2b**: will do so more strongly for individuals of higher ages and **H2c**: will do so more strongly for individuals with lower educational attainments.

Following our main hypotheses, we will explore the effects of our interventions and interaction with age and education on such precursors of acceptance as trust, expected safety and expected ease of use, to gain insights into which of these factors better explain participants' choices.

## Methods

### Data collection and participants

After obtaining ethics approval by the project ethics advisory board, we carried out the randomized online survey experiment on SoSciSurvey. Participants were German-speaking residents recruited via the research panel provider Bilendi and compensated in line with their renumeration scheme. We obtained a sample approximating the German population distribution on gender, age and education. G*Power calculations show that for our multiple linear regression model and a small effect size (partial $R^2 = 0.02$), 90% power could be achieved with N=813 participants. We over-recruited at N= 842. The average time spent on the survey (excluding the 3-minute video simulation) was 6 minutes (SD = 2:30; min: 1:25, max: 18). We excluded 43 speeders based on time values below the 0.05 quantile, i.e., the value below which 5% of the observations fall (participants completing the survey in less than 2.40 minutes), leaving us with a final sample of N=799 (achieved power for partial $R^2 = 0.02$ was 89.43%).

Of the 799 participants, N=423 self-reported as women (52.95%); N=342 were over the age of 50 (18-29: 18.27%; 30-39: 18.89%; 40-49: 20.02%; 50-59: 20.27%; 60-100: 22.52%). Educational levels were divided between participants that had completed schools at middle school or below (N=345, 43.17%), high school or equivalent (N=200, 25.03%), and university (N=254, 31.78%). Crossed sample demographics are presented in Figure 1 below.

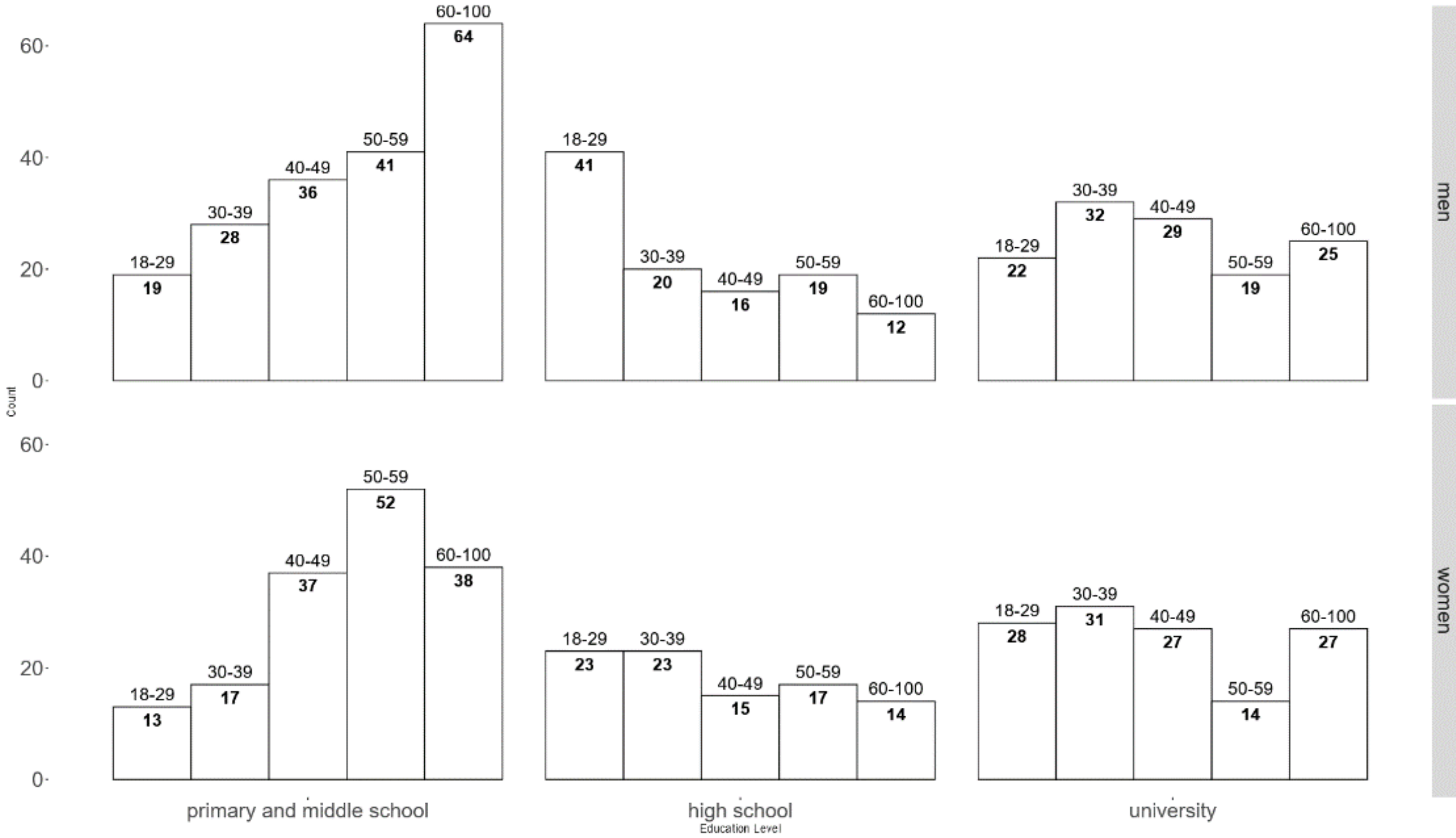

*Figure 1. Distribution of demographics, crossed gender, age and education.*

## Survey design and scales

The survey started with a letter of information, consent form, and questions about demographics (age, gender, education). After this, participants were given a brief introduction to Level 5 CAVs: "*Autonomous cars are slowly being introduced on our roads. These cars are not controlled by a human driver. Instead, they are completely controlled by a computer system. A fully autonomous car can take over all tasks and automatically controls all actions, including steering, acceleration, and braking. In the following survey, we are speaking of fully autonomous cars that require very little to no assistance from a human driver.*".

Participants were then randomized into intervention groups with the two factors (video simulation vs control, customization vs control), which will be explained in-depth below. After the intervention, participants were asked about their behavioral intention to use CAVs (Kacperski et al., 2021; J. Lee et al., 2019), their trust in CAVs (Choi & Ji, 2015), ease of use of CAVs (J. Lee et al., 2019), and expected consequences for safety (Kacperski et al., 2021). As a control characteristic, we queried participants' tech affinity (T. Franke et al., 2019). All these are self-report measures. Table 1 below showcases each scale, the associated Cronbach's alpha, mean and standard deviation. All scales were 7-point Likert scales ranging from "disagree completely" to "agree completely" unless otherwise noted. Scales were all averaged to a mean value for inclusion in regression models.

*Table 1. Overview over all scales and items. Columns list scale name, items, and the Cronbach's alpha, means and standard deviations of each scale. (r) marks reverse worsed items.*

| Scale | Items | Alpha | Mean | SD |
|---|---|---|---|---|
| Intention to use CAVs | If autonomous cars were available, I would use them.<br>I would be willing to use autonomous cars.<br>I would try to avoid autonomous cars as much as possible. (r)<br>I would not like to use autonomous cars. (r) | 0.95 | 3.75 | 2.02 |
| Trust | Overall, I think autonomous cars are dependable.<br>Overall, I think autonomous cars are reliable.<br>Overall, I can trust autonomous cars. | 0.97 | 3.84 | 1.76 |
| Safety | If large sections of the population used autonomous cars, travel for all citizens would be less safe (1) – safer (7).<br>If large sections of the population used autonomous cars, the number of traffic accidents would be higher (1) – lower (7) | 0.92 | 4.5 | 1.66 |
| Ease of use | I imagine that I would have problems using autonomous cars. (r)<br>I think I could handle autonomous cars well. | 0.83 | 3.91 | 1.78 |
| Tech affinity | I try to make full use of the capabilities of a technical system.<br>When I have a new technical system in front of me, I try it out intensively.<br>It is enough for me to know the basic functions of a technical system. (r)<br>I predominantly deal with technical systems because I have to. (r) | 0.71 | 4.33 | 1.31 |

## Interventions

Participants were randomized with SoSciSurvey's randomization algorithm into the four intervention groups. Two different interventions were either present or absent for participants:

(1) the video simulation, which included participants experiencing a video-simulated drive through a town (vs a control group that only received an image of the simulator interface, and was then forwarded to the survey questions). (2) Information about the opportunity to customize a CAV interface with regards to information (vs a control group with no customization).

**Video simulation**

If participants were randomized into the video-sim intervention, they were given the following instructions: "*You will now experience an autonomous car ride by watching a video simulation sequence of a ride in an autonomous vehicle. Put yourself in the following situation: You will start at home and drive to a shopping mall. The trip will take about 5 minutes. Please do not switch tabs/windows. Afterwards, there is an attention test that asks what you have seen.*"

On the following page, they were instructed: "*Turn on the sound on your computer. If you still have music/Youtube/etc. running in the background, please pause it for the duration of the video. Enable full screen mode in your browser by clicking the F11 key or the corresponding icon in your browser.*

*Did you follow all the instructions? Sound on? Full screen?*

*If you click "Next" now, you will be taken directly to the video simulation. When you are finished, you will be automatically redirected back to the questionnaire."*

Figure 2 below illustrates in more detail what participants saw in the video-sim, including multiple close-ups of interface details. It included a semi-realistic navigation interface with speed feedback, map view as well as navigation information in text, additional information about autonomous decision-making, and a voice navigation system that gave navigation information. Participants experienced five sections of a drive, including starting from parking position, overtaking at a construction site, a lane merge, a highway segment with exit sequence (due to traffic jam ahead), and parking at the destination.

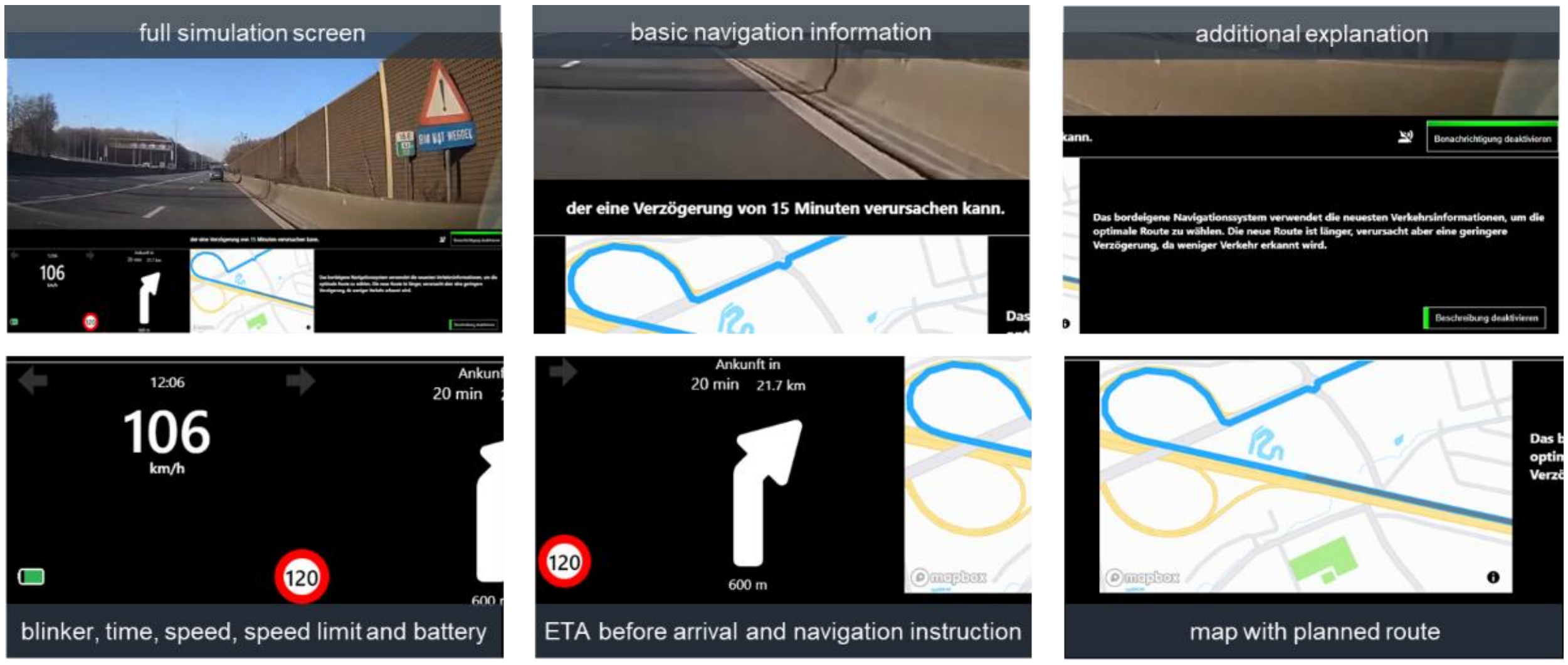

*Figure 2.* *Screenshots of the video simulation interface experienced by participants. Description of each screen is available above or below each of the screenshots.*

Following the video-sim, participants were asked three single-choice yes/no attention questions (example item: “In the driving sequences shown before, did your vehicle slow down to drive past a construction site?”). Participants who failed any were redirected back to the panel provider. Participants in the control group saw a still of the autonomous vehicle interface as an example of a possible autonomous vehicle interface (see Figure 3).

**Customization intervention**

Information customization in the car was introduced via the screenshots of the video-sim (Figure 3). Customization in this case meant that participants would have the option to choose how much information they wished to receive from the autonomous vehicle. Two types of information were available: navigation information (narrating information about the drive), for example “*A detour is being initiated, we will now exit the highway*.” and additional but non-essential background information to explain certain decision-making: "*A traffic jam has been detected on the current route, which may cause a delay of 15 minutes. The in-vehicle navigation system uses the latest traffic information to select the optimal route. The new route is longer but causes less delay because less traffic is detected*".

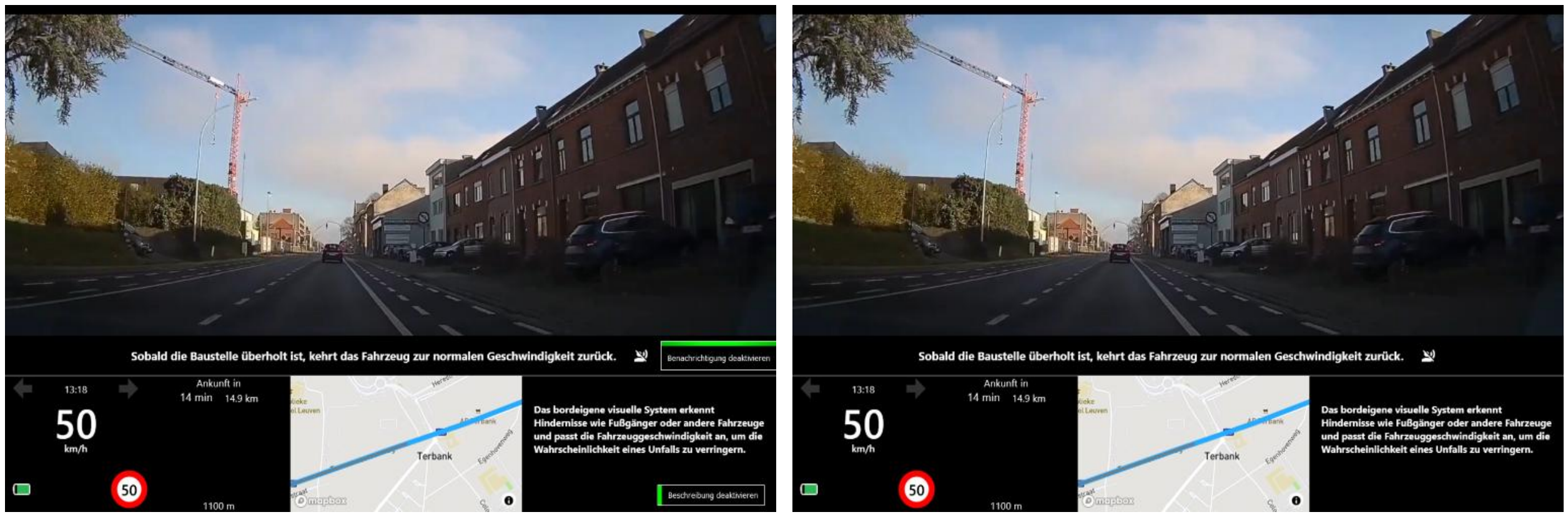

*Figure 3. Screenshot of the video simulation presented to participants, left-hand side: with customization buttons (green outlined buttons), right-hand side: control group with all information, no customization.*

In the condition with customization, participants' attention was drawn to their opportunity to customize how much information they might receive via the following item, to better activate the concept: "*As you can see on the picture above, you could customize information about the car's decisions to your needs (for example disable information you don't need). Please indicate your attitude towards this*." (not at all helpful – very helpful, 7-point scale). On average, participants in the customization condition rated this feature to be moderately helpful (M=5.23, SD=1.9). Participants in the control group received the right-hand image, which did not show the customization buttons, and they did not receive this item.

Participants assigned to the video-sim intervention group could interact with the illustrated customization buttons in the video-sim, i.e., they were able to toggle on and off information, and customize for themselves the amount of information they received from the navigation assistant. By default, this information was "on" for all video-sim intervention participants.

## Models

Data analysis was conducted with R statistics. We used the lm function of the lme4 package (Bates et al., 2015) to calculate linear regression models. We predicted our response variables (intention to use; trust; ease of use; safety expectations) from our interventions (video simulation and customization[1]), in interaction with age and education (ordered factor). We additionally controlled for gender as well as technological affinity, both of which have previously been found to impact CAV acceptance (Golbabaei et al., 2020; Nordhoff, Kyriakidis, et al., 2019). All estimates (B) are reported unstandardized, p-values at the 5% level are reported for significance testing. Interaction plots were created with interact_plot and cat_plot functions from the interaction package with 95% confidence intervals (Long, 2019).

# Results

## Variable correlations

Means and SDs of our scales can be found in Table 1 in the Methods section in the overview over the scales. Table 2 shows the correlation table. Our response variables (BI, trust, ease of use and safety expectations) were all correlated. Age and education were correlated with all response variables (age negatively, education positively), and slightly negatively correlated, i.e. older participants were less educated. Tech affinity was also positively correlated with all response variables. Being male was also associated with higher intention, safety, and ease of use expectations, but not trust.

---

[1] We explored the interaction between video simulation and customization but did not find significant effects of this on our response variables. We thus removed this interaction from our models.

*Table 2. Correlation table of all constructs.*

| | BI | trust | safety | ease of use | education | age | gender | tech affinity |
|---|---|---|---|---|---|---|---|---|
| BI | | 0.79* | 0.66* | 0.74* | 0.21* | -0.2* | 0.08* | 0.29* |
| trust | 0.79* | | 0.68* | 0.68* | 0.17* | -0.16* | 0.05 | 0.19* |
| safety | 0.66* | 0.68* | | 0.61* | 0.18* | -0.1** | 0.14* | 0.15* |
| ease | 0.74* | 0.68* | 0.61* | | 0.16* | -0.19* | 0.17* | 0.36* |
| education | 0.21* | 0.17* | 0.18* | 0.16* | | -0.19* | 0.04 | 0.06 |
| age | -0.2* | -0.16* | -0.1** | -0.19* | -0.19* | | 0.01 | -0.08* |
| gender | 0.08* | 0.05 | 0.14* | 0.17* | 0.04 | 0.01 | | 0.23* |
| tech affinity | 0.29* | 0.19* | 0.15* | 0.36* | 0.06 | -0.08* | 0.23* | |

*Note: * signifies $p < 0.05$. Education was coded as 1=middle school, 2=high, 3=university. Gender was coded as 1=women, 2=men. Response variables are on a 7-point scale.*

### Model outputs

Regression models can be found in Table 3. In line with H1a, experiencing the video simulation increased BI significantly. This finding repeats across all other response variables, with trust being affected highest after BI. In line with H1b, we found that there was a significant interaction with age, in that older participants' BI increased when they experienced the video simulation. We only found the same effect on the response variable ease of use, but not trust or safety expectations. We did not, counter to H1c, find a significant interaction with education.

Regarding, H2a, we did not find that the customization option affected BI significantly as a main effect. Regarding H2b, we found that older participants reported a lower intention to use if they were made aware of the customization option, and their reported safety expectations also decreased, while conversely, it was younger participants that increased in BI. Regarding H2c, participants with lower educational attainments reported higher intention to use in this case, while participants with higher educations decreased in BI. This pattern was also found for trust. Interaction plots related to BI can be seen in Figure 4.

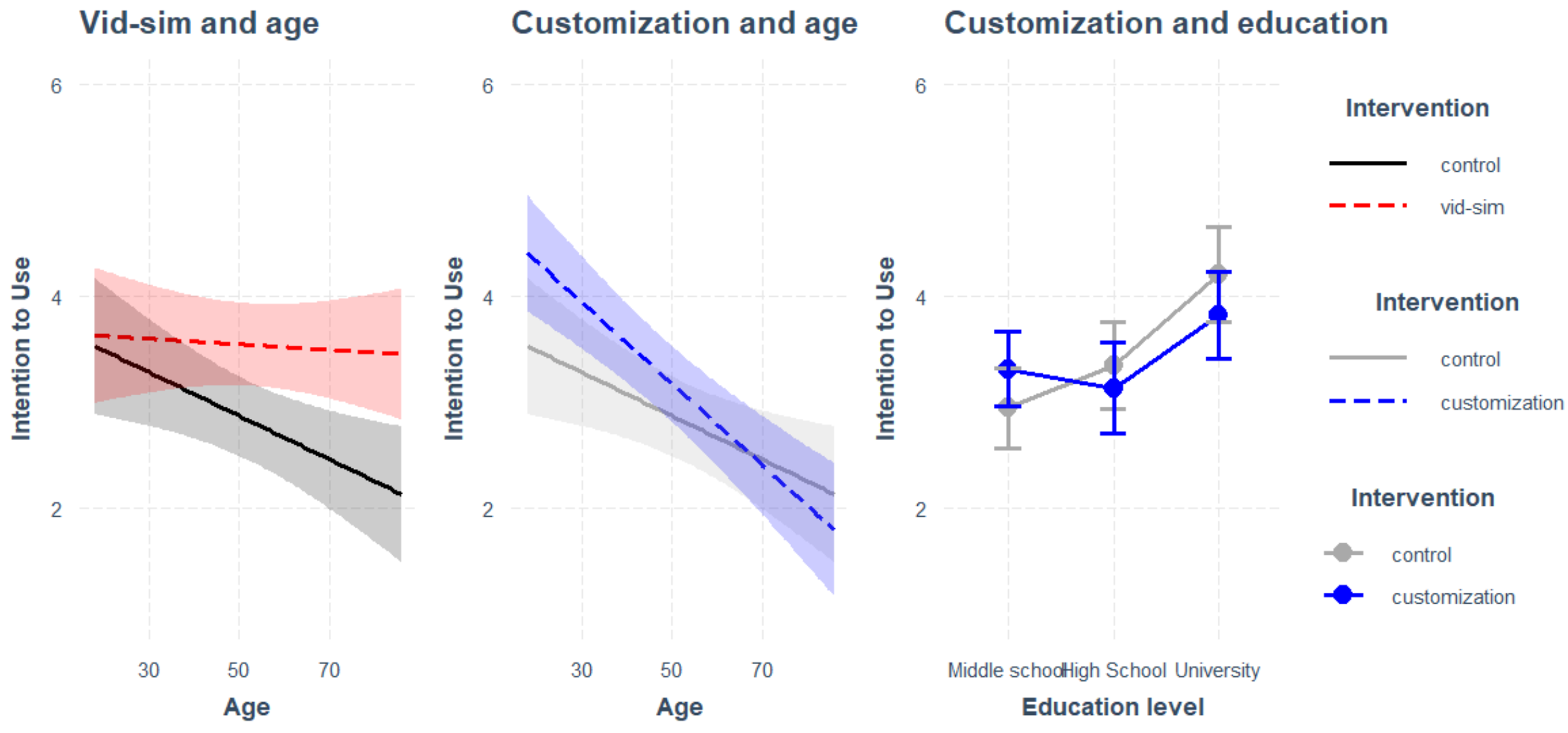

*Figure 4. Graphs of interaction of video simulation (vid-sim) intervention with age (left, red) and interaction of customization intervention with age (middle, blue) and educational level (right, blue).*

As a main effect, lower age and higher educational attainment was consistently associated with significantly more positive acceptance of CAVs. For our control variables, higher tech affinity was similarly consistently associated with more positive evaluation of CAVs. We only found gender effects on safety expectations and ease of use, with men having more positive expectations.

_Table 3. Regression results for models. Significant values are marked as **<0.01; * <0.05; † = 0.055. Significant interaction effects are bolded._

| | intention to use | | trust | | ease of use | | safety | |
|---|---|---|---|---|---|---|---|---|
| **predictor** | **B** | **95% CI** | **B** | **95% CI** | **B** | **95% CI** | **B** | **95% CI** |
| (intercept) | 3.50** | [3.24, 3.76] | 3.64** | [3.40, 3.87] | 3.62** | [3.39, 3.84] | 4.24** | [4.02, 4.47] |
| vid-sim | 0.59** | [0.33, 0.86] | 0.45** | [0.21, 0.69] | 0.35** | [0.12, 0.58] | 0.33** | [0.10, 0.56] |
| customization | -0.08 | [-0.34, 0.19] | -0.02 | [-0.26, 0.22] | -0.07 | [-0.30, 0.16] | -0.11 | [-0.34, 0.12] |
| education | 0.89** | [0.51, 1.27] | 0.66** | [0.32, 1.01] | 0.41* | [0.08, 0.74] | 0.47** | [0.14, 0.80] |
| age | -0.32** | [-0.56, -0.08] | -0.27* | [-0.49, -0.06] | -0.36** | [-0.57, -0.15] | -0.11 | [-0.31, 0.10] |
| tech affinity | 0.53** | [0.40, 0.66] | 0.30** | [0.18, 0.42] | 0.56** | [0.45, 0.68] | 0.20** | [0.09, 0.32] |
| gender | 0.08 | [-0.19, 0.34] | 0.02 | [-0.22, 0.26] | 0.34** | [0.11, 0.57] | 0.34** | [0.11, 0.57] |
| vid-sim * education | -0.12 | [-0.56, 0.31] | -0.13 | [-0.52, 0.26] | 0.18 | [-0.20, 0.56] | 0.08 | [-0.30, 0.45] |
| vid-sim * age | **0.28*** | [0.01, 0.55] | 0.19 | [-0.06, 0.43] | **0.25*** | [0.01, 0.48] | 0.17 | [-0.06, 0.40] |
| custom * education | **-0.54*** | [-0.97, -0.10] | **-0.40*** | [-0.79, -0.01] | -0.36 | [-0.74, 0.02] | -0.14 | [-0.52, 0.23] |
| custom * age | **-0.28*** | [-0.55, -0.01] | -0.13 | [-0.38, 0.11] | -0.11 | [-0.35, 0.12] | **-0.23†** | [-0.46, 0.01] |
| model fit $R^2$ | .179** | [.12,.21] | .106** | [.06,.13] | .198** | [.14,.23] | .089** | [.04,.11] |

## Discussion

We conducted an experimental study on the causal effect of experiencing a short video simulation with customization option of an autonomous vehicle drive on acceptance. We found that experiencing the video simulation positively affected CAV acceptance across our response variables intention to use, trust, and expectations of ease of use and safety. The experience of the video simulation more positively affected older participants' intention to use CAVs and their judgements of ease of use, but we did not find it to affect participants of varied educational backgrounds differently.

Our findings are in line with the literature on in vivo experiences of CAVs (Classen et al., 2021; Dai et al., 2021; Feys et al., 2021; Nordhoff et al., 2020). But while in vivo trials and tests can be time and cost intensive and cannot realistically be supplied to large numbers of citizens, our trial confirms that even experiencing a simple 3-minute video simulation can already affect trust and safety expectations, including intention to use CAVs, and can be useful as a research intervention to test other features of interest about CAVs, in line with previous arguments made comparing simulative and in vivo environments (Classen et al., 2020, 2021). We contribute to the growing literature showcasing simulations' effect on CAV acceptance (Clement et al., 2022; Park et al., 2019; Zontone et al., 2020), but extend such research generally conducted studying pre-post measurements by providing causal experimental evidence from comparison to a control group.

We particularly highlight exposure of older participants and positive experiences they may gain from interacting with such simulative environments and learning more about CAVs in this manner, which is of importance as older individuals have previously been less willing to accept CAVs for safety and usability reasons (Hassan et al., 2021; Shi et al., 2021; S. Wang & Zhao, 2019). Of interest is the fact that of our exploratory variables, judgements of ease of use from older participants was the only affected by the video simulation intervention – further research should investigate whether particularly ease of use is best targeted by simulations to encourage CAV usage.

Beyond the video simulation alone, we also investigated whether the opportunity to customize the navigation information interface would affect acceptance. While there is evidence for positive effects of letting drivers customize the interface and the amount of information their receive (DeGuzman et al., 2022; S. Lee et al., 2020; Normark, 2015; Wintersberger et al., 2019), in our case, we found, contrary to our hypothesis, that customization positively affected CAV acceptance for younger individuals, while it lowered older individuals' intention to use. Apparently, older adults were not overwhelmed by the customization features as effects on ease of use and trust were not significant. Instead, the freedom to choose decreased their belief in the safety of CAVs. For individuals with lower education levels, customization positively affected intention to use CAVs and trust, while it lowered them for university educated participants. More research is needed in this direction, for example, future research might be able to clarify whether this is more related to cognitive load in some way, or related to feelings of trust or safety, which were the response variables on which we also found effects.

**Limitations**

We report data from a panel sample of German citizens, which, while approximating the German population, is not fully representative, generalizations should therefore be made with care. While we included participants of all age groups and educational strata, further research could, when studying age effects, more specifically focus on large samples of participants above sixty years of age as a target group, or those with lower educational attainments.

While other variables such as gender and income have also been noted to be associated with CAV acceptance (Charness et al., 2018; Golbabaei et al., 2020), we felt including a greater number of socio-demographics in our interaction analyses would take away from the narrow focus we wanted to achieve; our selection of age and education as primary predictors was driven by their consistent prominence in the literature and their alignment with the overarching goal of promoting equity and inclusivity in technology adoption. These groups are also likely to benefit significantly from autonomous vehicle technology if it is implemented inclusively, making them critical populations to address. Future research could in-depth investigate and compare whether gender differences in acceptance are driven by other mechanisms than those of age and education, and if simulation and customization interventions would need to be adapted to make CAVs more attractive for women.

While we consider it an asset that we were able to improve on acceptance with a very simple, brief video simulation intervention, it should be pointed out that in reality, wide-spread deployment of Level 5 navigation as presented here is still in the future (Khan et al., 2022) and might look very differently from the way we presented it to our audience, as the concrete implementation is yet unclear. Testing more variants is important to understand whether they might lead to different levels of acceptance. Similarly, our customization intervention was kept minimalist and only provided participants with the option to pick and choose information in a very narrow context. Future studies could expand the customization options and further study its limitations, as customization and personalization might fast become overwhelming to a detrimental effect for acceptance (S. Lee et al., 2020), especially considering differing information processing requirements and skills (Ackermann, 2003; Phillips & Sternthal, 1977).

Finally, as our study consists of an online experiment, our response variables measure self-reported acceptance, and while the experimental design allows for causal conclusions in the context that the study was carried out, generalizations onto acceptance in realistic contexts should be made with care: any CAV simulator has the drawback that participants might fail to realize distinctions to real-world scenarios.

**Policy implications**

Existing policies are not currently designed or updated for transport infrastructures that include CAVs (D. Lee & Hess, 2020). Yet, the potential of CAVs to address existing transportation inequities hinges on proactive policy interventions. Marginalized groups, such as low-income, lower educated communities, older adults, and rural residents, risk being left behind in the absence of equitable deployment strategies: already, transport justice frameworks advocate for a redistribution of transport benefits and burdens (Martínez-Buelvas et al., 2022). Novel policies must address a diverse range of issues, including safety, inclusivity, equity, and public trust, while anticipating the long-term impacts (Schepis et al., 2023). Governments will here play a dual role as facilitators of innovation and enforcers of regulations - beyond traditional regulatory functions, public agencies can actively shape CAV markets by funding research, piloting public CAV services, and incentivizing sustainable usage models (Schepis et al., 2023).

Our study's support for the familiarity principle, which posits that mere exposure improves acceptance (Bornstein, 1989; Montoya et al., 2017), suggests that brief, cost-effective video simulations can serve as scalable, effective interventions to mitigate safety concerns and improve ease of use, as well as practical tools for researchers and policymakers to investigate mechanisms across broader populations and pilot novel interactive solutions. For example, in our study, we find some evidence that excessive customization may heighten perceptions of system unreliability for some populations (Grice, 1975; Beller et al., 2013) - practitioners could instead adopt a balanced approach to customization, providing simplified, pre-set options for older users while retaining flexibility for technologically experienced and younger demographics.

It is worth discussing whether regulations might be necessary regarding UI design, and whether licenses and educational campaigns should in the future shift towards training with more autonomous and UI-related features. Short, targeted training modules as part of driver's license trainings could familiarize users with the safe operation of CAV systems. This approach aligns with the broader goal of ensuring inclusivity by addressing potential technological literacy gaps. Optimally, practical demonstrations that target risk and usability through direct interaction with CAV systems could be incorporated. Additionally, regulatory bodies such as UNECE (UNECE, 2024) could consider introducing age-dependent interface requirements from manufacturers, whereby default displays minimize information overload but allow for advanced customization by licensed users. Participatory approaches could additionally ensure the technology addresses the unique needs of vulnerable populations such as those with visual impairments (Brinkley et al., 2017; Kacperski et al., 2024), fostering broader acceptance.

In summary, a potential full decoupling of human drivers from vehicle behavior necessitates among others a re-evaluation of licensing, liability, and user interface (UI) standards. The findings of this study underscore the importance of simulated CAV interventions on their own, but also the tailoring of such to the diverse needs of population subgroups, such as older individuals and those with lower educational attainment that we focus on in this study.

## Conclusion

We report findings from a study on acceptance of CAVs following a video simulation intervention and an intervention that introduces participants to customization of navigation information. We find that even a very simple video simulation can already improve acceptance across a variety of response variables such as intention to use, trust, ease of use, safety expectations, and that such interventions are particularly effective to improve acceptance among older individuals. An intervention about the opportunity to customize navigation information might negatively affect older individuals and those with university degrees, but improve acceptance by younger individuals and those of lower educational levels.